\documentclass[conference,a4paper,twocolumn]{IEEEtran}

\IEEEoverridecommandlockouts

\ifCLASSINFOpdf
\else
\fi
\usepackage[cmex10]{amsmath}
\usepackage[ruled,vlined]{algorithm2e}

\usepackage{stfloats}
\usepackage{url}
\usepackage{graphicx}
\usepackage{amssymb}
\usepackage{hyperref}
\usepackage{xcolor}
\usepackage[T1]{fontenc}
\usepackage[utf8]{inputenc}
\usepackage{multirow}
\usepackage{subcaption}

	\usepackage{pgfplots}
  \pgfplotsset{compat=newest}
\newlength\figurewidth
\usepackage[skip=4pt]{caption}
\def\RR{\mathbb{R}}

\def\CC{\mathbb{C}}

\def\x{\vect{x}}
\def\u{\vect{u}}
\def\vv{\vect{v}}
\def\y{\vect{y}}
\def\z{\vect{z}}

\def\p{\vect{p}}
\def\q{\vect{q}}
\def\c{\vect{c}}

\def\xq{\vect{x}^\mathrm{q}}

\def\D{\matr{D}}

\newcommand{\norm}[1]{\|#1\|}

\newcommand{\vect}[1]{\mathbf{#1}} %vector
\newcommand{\matr}[1]{\mathbf{#1}} %matrix

\newcommand{\argmin}{\mathop{\operatorname{arg~min}}}

\newcommand{\soft}{\mathrm{soft}}
\newcommand{\proj}{\mathrm{proj}}

\newcommand{\sgn}{\mathrm{sgn}}
\newcommand{\clip}{\mathrm{clip}}

\newcommand{\syn}{\ensuremath{D}}
\newcommand{\ana}{\ensuremath{A}}

\newcommand{\dsdr}{\ensuremath{\Delta\text{SDR}}}

\newcommand{\qm}[1]{``#1''}  %quotation marks

\begin{document}

% ---------------------------------------------------
% Title
% ---------------------------------------------------


\title{Sparse and Cosparse Audio Dequantization \\Using Convex Optimization}
% Audio Signal Dequantization Based on Signal Sparsity
% ---------------------------------------------------
% Authors, Affiliation, and Acknowledgment
% ---------------------------------------------------

\author{\IEEEauthorblockN{
Pavel Záviška, Pavel Rajmic
}
\IEEEauthorblockA{
Signal Processing Laboratory\\
Brno University of Technology\\
Brno, Czech Republic\\
Email: xzavis01@vutbr.cz, rajmic@vutbr.cz
}
\thanks{The work was supported by the project 20-29009S of the Czech Science Foundation (GAČR)
and by the joint project of the FWF and the Czech Science Foundation: numbers I3067-N30 and 17-33798L, respectively.
}}


%Telephone: (800) 555--1212, Fax: (888) 555--1212}


\maketitle

% ---------------------------------------------------
% Abstract
% ---------------------------------------------------

\begin{abstract}
%The abstract goes here. The length of the abstract should not exceed 150 words.
%
The paper shows the potential of sparsity-based methods in restoring quantized signals.
Following up on the study of Brauer et al.\ (IEEE ICASSP 2016),
we significantly extend the range of the evaluation scenarios:
we introduce the analysis (cosparse) model, we use more effective algorithms, 
we experiment with another time-frequency transform.
The paper shows that the analysis-based model performs comparably to the synthesis-model,
but the Gabor transform produces better results than the originally used cosine transform.
Last but not least, we provide codes and data in a~reproducible way.
\end{abstract}

% ---------------------------------------------------
% Keywords
% ---------------------------------------------------


% \fi
\IEEEpeerreviewmaketitle

% ---------------------------------------------------
% Introduction
% ---------------------------------------------------

\section{Introduction}
%\IEEEPARstart
Signal quantization is inherently present in all areas of digital signal processing (DSP).
The need for quantization arises from the fact that signals in DSP have to be stored and processed using finite arithmetic.
Quantization appears in the digitization of analog signals, in the field of generating artificial signals,
and it is widely used in various compression algorithms
\cite{VaryMartin2006:DST,ITU-T_G.711,Zolzer2011:DAFX,PearlmanSaid2011:Digital.signal.compression}.

A quantized signal is always distorted by quantization, and such a~nonlinear operation is not reversible in general.
The process of estimating the original signal from its quantized counterpart is usually called dequantization or
bit depth expansion and it overlaps with problems within the area of decompressing signals.
Since dequantization is an ill-posed inverse problem, dequantization methods must rely on some additional information.
Most frequently, the mathematical or statistical properties of the (unknown) original signal are formulated;
these properties are typically violated by the quantized signal, making it possible to formulate an optimization problem
that prioritizes desirable signals
and whose solution leads to a~dequantized signal.

Within the field of audio processing, paper
\cite{Troughton1999:Dequantization.Sinusoidal.AR} utilizes a~sinusoidal model
on the assumption that the model residuals are autoregressive random processes.
The paper reports an average distortion reduction by 11\,dB.

Methods based on signal sparsity \cite{Bruckstein.etc.2009.SIAMReviewArticle} have become most popular in the last decade.
Paper 
\cite{BrauerGerkmannLorenz2016:Sparse.reconstruction.of.quantized.speech.signals}
designs a simple sparsity-based model and evaluates it on a~speech database.
This paper will serve us as a~reference.

Dequantization and declipping are treated at once in the recent contribution \cite{RenckerBachWangPlumbley2018:Fast.iterative.shrinkage.declip.dequant-iTwist18}.
Note that the unified model for both the problems is natural since declipping
(i.e., signal amplitude saturation) can be considered a special case of quantization. 
Different sparsity-based computational methods are compared here, but only on artificially created signals,
and with no comparison to another existing method.

The same authors later extended this approach and combined it with learning the sparsifying transform
so as to make the signal restoration quality as high as possible, obviously traded-off with computational cost
\cite{RenckerBachWangPlumbley2018:Sparse.recovery.dictionary.learning}.

In the context of compressed sensing \cite{candes_cs}, dequantization has been discussed 
in \cite{Moshtaghpour2016:C-BP.for.Signal.and.Matrix.Estimates.in.Quantized.CS}.
Here, the difference is that quantized are not directly the signal samples as usual,
but their values after a~certain linear transformation.

Finally, the recent contribution 
\cite{BrauerZhaoLorenzFingscheidt2019:Dequantization_speech_signals}
follows up on
\cite{BrauerGerkmannLorenz2016:Sparse.reconstruction.of.quantized.speech.signals},
which has been discussed above, and modifies the original dequantization algorithm
such that it involves a~neural network learned on speech signals.

Our paper was motivated by promising results in \cite{BrauerGerkmannLorenz2016:Sparse.reconstruction.of.quantized.speech.signals}.
Nevertheless, \cite{BrauerGerkmannLorenz2016:Sparse.reconstruction.of.quantized.speech.signals}
is just a brief conference contribution with limited information value.
Therefore, our goal was to verify the results presented (there are no codes available),
use more effective algorithms, and significantly expand the evaluation scenarios.
Our paper thus uses a simpler Douglas--Rachford algorithm, where applicable;
we also include the so-called analysis (cosparse) signal model;
we add another time-frequency transform;
we do not limit ourselves to real-time applications and let the iterative methods fully converge.

Note that we do not compare our results either to 
\cite{RenckerBachWangPlumbley2018:Fast.iterative.shrinkage.declip.dequant-iTwist18}
or 
\cite{RenckerBachWangPlumbley2018:Sparse.recovery.dictionary.learning},
since the nature of their approach allows the so-called \emph{inconsistent} solutions,
meaning that it can happen that the quantization of the dequantized sample does not match the input quantized sample.
Such an approach could be justified  when the signal is noisy,
but this does not fall within our setup and the setup of
\cite{BrauerGerkmannLorenz2016:Sparse.reconstruction.of.quantized.speech.signals},
where the reconstructions are \emph{consistent}.

\begin{figure*}
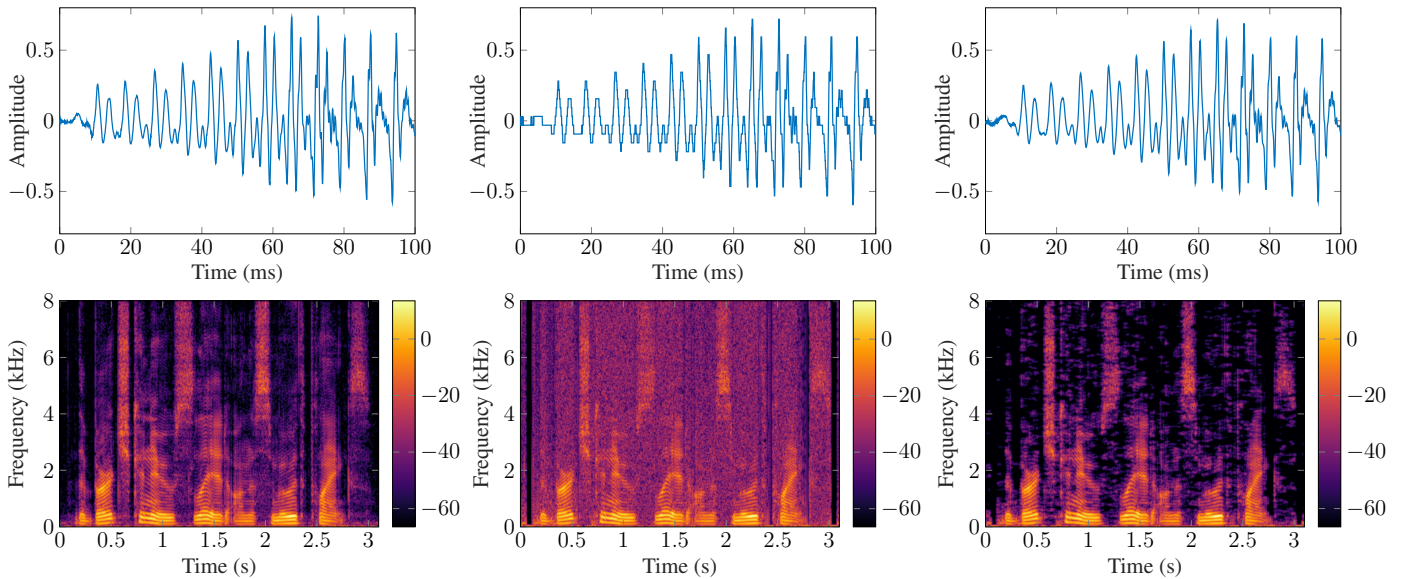
%
	\begin{subfigure}[b]{0.32\textwidth}\label{subfig:waveform_orig}
\definecolor{mycolor1}{rgb}{0.00000,0.44700,0.74100}%
\begin{tikzpicture}[scale=0.59]

\begin{axis}[%
width=3.12in,%5.179in,
height=2in,%3.842in,
at={(0.535in,0.444in)},
scale only axis,
xmin=0,
xmax=100,
xlabel style={font=\Large\color{white!15!black}},
xlabel={Time (ms)},
xticklabel style={font=\Large},
ymin=-0.8,
ymax=0.8,
ylabel style={font=\Large\color{white!15!black}},
ylabel={Amplitude},
ylabel shift=-20pt,
yticklabel style={font=\Large},
axis background/.style={fill=white},
every axis plot/.append style={thick}
]
\addplot [color=mycolor1] table {figz/waveform_orig.dat};
\end{axis}
\end{tikzpicture}%
	\end{subfigure} \hspace*{0.35pt}
	\begin{subfigure}[b]{0.32\textwidth}\label{subfig:waveform_quantized}
\definecolor{mycolor1}{rgb}{0.00000,0.44700,0.74100}%
\begin{tikzpicture}[scale=0.59]

\begin{axis}[%
width=3.12in,%5.179in,
height=2in,%3.842in,
at={(0.535in,0.444in)},
scale only axis,
xmin=0,
xmax=100,
xlabel style={font=\Large\color{white!15!black}},
xlabel={Time (ms)},
xticklabel style={font=\Large},
ymin=-0.8,
ymax=0.8,
ylabel style={font=\Large\color{white!15!black}},
ylabel={Amplitude},
ylabel shift=-20pt,
yticklabel style={font=\Large},
axis background/.style={fill=white},
every axis plot/.append style={thick}
]
\addplot [color=mycolor1] table {figz/waveform_quant.dat};
\end{axis}
\end{tikzpicture}%
	\end{subfigure} \hspace*{2pt}
	\begin{subfigure}[b]{0.32\textwidth}\label{subfig:waveform_rec}
\definecolor{mycolor1}{rgb}{0.00000,0.44700,0.74100}%
\begin{tikzpicture}[scale=0.59]

\begin{axis}[%
width=3.12in,%5.179in,
height=2in,%3.842in,
at={(0.535in,0.444in)},
scale only axis,
xmin=0,
xmax=100,
xlabel style={font=\Large\color{white!15!black}},
xlabel={Time (ms)},
xticklabel style={font=\Large},
ymin=-0.8,
ymax=0.8,
ylabel style={font=\Large\color{white!15!black}},
ylabel={Amplitude},
ylabel shift=-20pt,
yticklabel style={font=\Large},
axis background/.style={fill=white},
every axis plot/.append style={thick}
]
\addplot [color=mycolor1] table {figz/waveform_rec.dat};
\end{axis}
\end{tikzpicture}%
	\end{subfigure}\\ 
	%%%%%%%%%%%%%%%%%%%%%%%%%%%%%
	%SPECTROGRAMZ
  %%%%%%%%%%%%%%%%%%%%%%%%%%%%%
	\begin{subfigure}[b]{0.32\textwidth}\label{subfig:spectrogram_orig}
			\input{figz/Spectrogram_orig}
	\end{subfigure} \hspace{2pt}
	\begin{subfigure}[b]{0.32\textwidth}\label{subfig:spectrogram_quantized}
			\input{figz/Spectrogram_quant}
	\end{subfigure} \hspace{2pt}
	\begin{subfigure}[b]{0.32\textwidth}\label{subfig:spectrogram_rec}
			\input{figz/Spectrogram_rec}
	\end{subfigure}
\caption{Waveforms (top) and spectrograms (bottom) of the original (left), quantized (middle) and restored (right) signals.
Here, the spectrograms are generated from the whole utterance \qm{S\_01\_01.wav} sampled at 16\,kHz.
The waveforms are snippets 100~ms long taken from the middle of the utterance.
The word length of the quantized sample is $w=5$.
}
\label{fig:Waveforms_and_Spectrograms}%
\end{figure*}

We recall the signal quantization process in Sec.\,\ref{Sec:Quantization}. 
In Sec.\,\ref{sec:problem_formulation}, the dequantization problem is formalized,
and algorithms for its solution are proposed in Sec.\,\ref{Sec:Algorithms}.
Finally, Sec.\,\ref{Sec:Experiments} describes the experiments and evaluates the results.

% ---------------------------------------------------
% Quantization
% ---------------------------------------------------
\section{Quantization}
\label{Sec:Quantization}
Signal quantization is a nonlinear process of limiting the number of possible values the signal can attain.
Quantization typically arises as a~necessary step in the process of signal digitization, 
where each signal sample is rounded to the nearest quantization level.
Since digital signals are stored in binary form, the number of quantization levels is dependent on the amount of assigned bits, i.e., 
the word length \cite{Zolzer2011:DAFX},
which is given in bits per sample (bps). 

According to the placement of quantization levels,
it is possible to distinguish two different types of quantization: %---uniform and non-uniform quantization. 
In the case of uniform quantization, all the quantization levels are equally distributed over the whole dynamic range;
thus the quantization step, $\Delta$, is constant.
Obviously, this is the most naive approach to quantization. 
To minimize the overall quantization error, it is often beneficial to exploit the distribution of the values of samples and place the quantization levels non-uniformly across the whole dynamic range. 
Audio signals tend to concentrate the sample values around zero,
therefore it is advantageous to use a~nonlinear quantization scale \cite{VaryMartin2006:DST} 
like the A-law or $\mu$-law defined in the ITU-T Recommendation G.711 \cite{ITU-T_G.711}.

The present paper is devoted to the restoration of quantized signals.
For simplicity, we use the standard, so-called mid-riser uniform quantizer as the origin of the distortion in audio signals,
where the size of the quantization step is given by $\Delta = 2^{-w+1}$, with $w$ representing the word length in bps. 
The quantized signal $\xq \in \RR^N$ is computed according to the following formula:
\begin{equation}
(\xq)_n = \sgn^+ (\x_n) \Delta \left( \left\lfloor \frac{|\x_n|}{\Delta} \right\rfloor + \frac{1}{2} \right),
\label{eq:uniform_quantization}
\end{equation}
where $n$-th sample of the signal is denoted by the index $n$
and $\sgn^+(z)$ returns $1$ for $z\geq0$ and $-1$ for $z<0$.

Note, however, that the methods presented in this paper are completely independent of a particular choice of the quantizer.

{Waveforms and spectrograms of the original, quantized and restored signals
 are presented in Fig.\,\ref{fig:Waveforms_and_Spectrograms}.}
%\todo{ref to Fig.\,1 here?}

% ---------------------------------------------------
% Problem Formulation
% ---------------------------------------------------
\section{Problem Formulation}
\label{sec:problem_formulation}
Based on the quantized observation, $\xq$,
dequantization aims at restoring the signal to be as close as possible to the original (unknown) signal $\x$. 
But without any additional information about the signal, this task would be ill-posed.
It is therefore crucial that additional knowledge is considered.
Based on the fact that audio signals have approximately sparse coefficients in a~suitable time-frequency representation, it is possible to 
formulate the restoration task as finding the signal with the sparsest coefficients
whose samples will not lie further than $\frac{\Delta}{2}$ from the respective quantization level.

\subsection{Synthesis Case}
In the synthesis case, we use the model which assumes that the signal is composed of a linear combination of atoms from the matrix $\D$, 
which is usually called the dictionary and can also be understood as the linear operator $D\colon \CC^P \to \RR^N$.
Formally, it is possible to write $\x = D\c$, where $\x\in\RR^N$ is the time domain signal
and $\c\in\CC^P$ is a~vector of coefficients (sparse or close to sparse). %representation of the vector $\x$.
{In this paper, the operator $D$
%and the later-introduced operator $A$
forms the so-called Parseval tight frame, which is advantageous for the derivation of the algorithms.}

Since finding the true sparsest vector is an NP-hard problem,
we use the so-called relaxation and replace the non-convex $\ell_0$ norm with the closest convex norm,
which is the $\ell_1$ norm.
The synthesis-based dequantization problem can be therefore formulated as finding
\begin{equation}
\argmin_\c \norm{\c}_1 \hspace{1em} \mathrm{ s.t. } \hspace{1em} \norm{{\syn\c - \xq}}_\infty < \frac{\Delta}{2},
\label{eq:synthesis_problem_const}
\end{equation}
where $\norm{\cdot}_\infty$ denotes the $\ell_\infty$ norm returning the largest magnitude of the input vector.

To solve \eqref{eq:synthesis_problem_const},
it is convenient to rewrite it into an unconstrained form:
\begin{equation}
\argmin_\c \norm{\c}_1 + \iota_{\Gamma^*}(\c),
\label{eq:synthesis_problem_unconst}
\end{equation}
where the constraint is replaced by the addition of the indicator function $\iota_{C}(\cdot)$,
which returns 0 if its argument lies in the convex set $C$, and $\infty$ otherwise.
In  \eqref{eq:synthesis_problem_const}, the convex set $\Gamma^*\subset\CC^P$ is
defined as
\begin{equation}
\Gamma^* = \{\c'\ |\ \norm{{\syn\c' - \xq}}_\infty < \frac{\Delta}{2}\}.
\label{eq:gamma_synth}
\end{equation}

\subsection{Analysis Case}
In contrast to the synthesis model, the dequantization problem can also be formulated as analysis-based,
often also called cosparse.
Here, instead of composing the signal from several components,
the analysis operator $A$ is used, which generates coefficients from the signal,
such that
$A\x = \c$, where $A\colon \RR^N\to\CC^P$ and it holds that $D^* = A$ (notation~$^*$~represents the adjoint operator). 

Using the cosparse model, the dequantization problem attains the form
\begin{equation}
\argmin_\x \norm{\ana\x}_1 \hspace{1em} \mathrm{ s.t. } \hspace{1em} \norm{{\x - \xq}}_\infty < \frac{\Delta}{2}.
\label{eq:analysis_problem_const}
\end{equation}
As with the synthesis case,
%\eqref{eq:synthesis_problem_unconst}
we rewrite the problem to the unconstrained form:
\begin{equation}
\argmin_\x \norm{\ana\x}_1 + \iota_\Gamma(\x),
\label{eq:analysis_problem_unconst}
\end{equation}
where the convex set of feasible solutions $\Gamma\subset\RR^N$ is now a~very simple set of time domain signals:
\begin{equation}
\Gamma = \{\x'\ |\ \norm{{\x' - \xq}}_\infty < \frac{\Delta}{2}\}.
\label{eq:gamma_analysis}
\end{equation}

% ---------------------------------------------------
% Algorithmic solution
% ---------------------------------------------------
\vspace{0.2em}
\section{Algorithmic Solution}
\label{Sec:Algorithms}
To solve problems defined in Sec.~\ref{sec:problem_formulation},
the proximal splitting algorithms are used \cite{combettes2011proximal}. 
The synthesis-based problem \eqref{eq:synthesis_problem_unconst} consists in minimizing a sum of two convex functions,
and in this case the Douglas--Rachford (DR) algorithm can be used % to \todo{quite efficiently???}.
\cite{combettes2007douglas}.
The algorithm is presented in Alg.\ \ref{alg:DR_dequantization}.
\vspace{-0.2em}
\begin{algorithm}[ht]%[H]
	\DontPrintSemicolon
	\SetAlgoVlined
	\KwIn{Set starting point $\c^{(0)} \in \CC^P.$ \\
				Set parameters $\lambda =1, \gamma > 0$.
	}
	\For{$i=0,1,\dots$\,}{
		$\tilde{\c}^{(i)} = \proj_{\Gamma^*}\c^{(i)}$ \;
		$\c^{(i+1)} = \c^{(i)} + \lambda \left(\soft_{\gamma}(2\tilde{\c}^{(i)} - \c^{(i)}) - \tilde{\c}^{(i)} \right)$ \;
	}
	\KwRet{$\c^{(i+1)}$}
	\caption{\mbox{Douglas--Rachford (DR) algorithm solving \eqref{eq:synthesis_problem_unconst}}}
	\label{alg:DR_dequantization}
\end{algorithm}%\DecMargin{1em}
\vspace{-0.2em}

The algorithm consists of two principal steps. 
The first step is the projection onto $\Gamma^*$. 
Using the projection lemma introduced in \cite{RajmicZaviskaVeselyMokry2019:Axioms},
the projection can be efficiently computed by
\begin{equation}
\proj_{\Gamma^*}(\z) = \z - \syn^*\left(\syn\z-\proj_{\Gamma}(\syn\z)\right),
\label{eq:projection}
\end{equation}
where the projection onto $\Gamma$ at the right side of \eqref{eq:projection} is a~trivial time-domain mapping:
\begin{equation}
\Big(\proj_\Gamma(\y)\Big)_n = \left\{\hspace{-0.2em}
	\begin{array}{ll}
		\y_n & \text{if } |\y_n - (\xq)_n| < \frac{\Delta}{2}, \vspace{0.3em}\\
		(\xq)_n + \frac{\Delta}{2} & \text{if } \y_n - (\xq)_n < - \frac{\Delta}{2}, \vspace{0.3em}\\
		(\xq)_n - \frac{\Delta}{2} & \text{if } \y_n - (\xq)_n > \frac{\Delta}{2}. 
	\end{array}
	\hspace{-0.2em}
	\right.
\label{eq:proj_time}
\end{equation}

The second principal step of the algorithm is soft thresholding as the proximal operator of the $\ell_1$ norm,
defined as
\begin{equation}
\soft_{\gamma}(\z) = \sgn(\z)\odot\max(|\z| - \gamma, 0),
\label{eq:soft_thr}
\end{equation}
where $\odot$ represents the elementwise product. 

In the analysis case, the linear operator $A$ inside the $\ell_1$ norm in \eqref{eq:analysis_problem_unconst}
disables us to use the DR algorithm as in the synthesis case. 
%Therefore, it is necessary to use an algorithm, that is capable of dealing with the linear operator inside one of the functions. 
%Such an algorithm is called Chambolle--Pock algorithm and its form suited for signal dequantization is shown in Alg.\ \ref{alg:CP_dequantization}.
The Chambolle--Pock (CP) algorithm \cite{ChambollePock2011:First-Order.Primal-Dual.Algorithm} is able to cope with a linear operator with one of the functions.
Its particular form for signal dequantization is shown in Alg.~\ref{alg:CP_dequantization}.
\vspace{-0.2em}
\begin{algorithm}%[H]
\DontPrintSemicolon
\SetAlgoVlined
\KwIn{Set starting points $\p^{(0)} \in \RR^N, \q^{(0)} \in \CC^P.$ \\
	    Set parameters $\zeta, \sigma > 0$ and $\rho\in[0,1]$.
}
\For{$i=0,1,\dots$\,}{ 
	$\textbf{q}^{(i+1)} = \mathrm{clip}_{1}(\textbf{q}^{(i)}+\sigma \ana \bar{\textbf{p}}^{(i)})$\\
	$\textbf{p}^{(i+1)} = \proj_{\Gamma}(\textbf{p}^{(i)}-\zeta \ana^*{\textbf{q}}^{(i+1)})$\\
	$\bar{\textbf{p}}^{(i+1)} = \textbf{p}^{(i+1)} + \rho (\textbf{p}^{(i+1)} - \textbf{p}^{(i)}) $\\
} 
\KwRet{\(\bar{\textbf{\textup{p}}}^{(i+1)}\)}
\caption{\mbox{Chambolle--Pock (CP) algorithm solving \eqref{eq:analysis_problem_unconst}}}
\label{alg:CP_dequantization}	
\end{algorithm}
\vspace{-0.2em}

The algorithm has again two principal steps corresponding to the proximal operators of the minimized functions.
Since the set of feasible solutions $\Gamma$ is defined in the time-domain, the projection step consists in the simple elementwise mapping as in \eqref{eq:proj_time}.
Because of the Fenchel--Rockafellar conjugate inside the first step of the algorithm, the soft thresholding becomes a clip function:
\begin{equation}
\clip_{\lambda}(\x) = \sgn(\x) \odot \min(|\x|, \lambda).
\label{eq:clip}
\end{equation}
For $\rho=1$, the algorithm is proved to converge if $\zeta\sigma\norm{A}^2<1$, where $\norm{\cdot}$ denotes the operator norm.

% ---------------------------------------------------
% Experiments and results
% ---------------------------------------------------
\section{Experiments and Results}
\label{Sec:Experiments}

The experiments were designed to evaluate the performance of sparsity-based dequantization methods,
using the $\ell_1$-minimization in both the synthesis and the analysis models.
%Different types of transform, \new{such as 
The Discrete Gabor Transform (DGT) and the Windowed Modified Discrete Cosine Transform (WMDCT) were included in the testing, %\todo{explain DGT and WMDCT here}
and two means of numerical evaluation were used, specifically the Signal-to-Distortion Ratio (SDR) and Perceptual Evaluation of Speech Quality (PESQ).
See details below.

The original intent of the paper was to extend the experiments from 
\cite{BrauerGerkmannLorenz2016:Sparse.reconstruction.of.quantized.speech.signals},
which were only limited to a single transform and to the synthesis model.
Unfortunately, even though we used the same audio database, the same quantization type, and the same evaluation methods,
 %(PESQL from \cite{Loizou2017:Speech.Enhancement}),
we found out that we were not able to reproduce the
%the same PESQL values for quantized signals.
results reported in \cite{BrauerGerkmannLorenz2016:Sparse.reconstruction.of.quantized.speech.signals}.
Therefore, our paper can be considered a~stand-alone study of dequantization algorithms based on sparsity.

The audio database, provided as an online archive attached to the book \cite{Loizou2017:Speech.Enhancement},
consists of 720 male speech utterances with an approximate length of 2 seconds.
These speech signals are available at the sampling frequencies 8\,kHz (narrowband) and 25\,kHz (wideband) in wav files with 16 bps. 
Since the PESQL evaluator is able to process only signals sampled at 8\,kHz or 16\,kHz,
the signals were first downsampled from 25 to 16\,kHz. % in the pre-processing step.

After the downsampling process, the speech signals were peak-normalized and then uniformly quantized according to the quantization rule in \eqref{eq:uniform_quantization}.
The signal was degraded to seven different grades, using the word lengths $w = 2, 3, \dots, 8$.

Assuming that the reconstructed signals have a sparse representation, 
the DR algorithm (Alg.\,\ref{alg:DR_dequantization}) is used to approximate the solution of \eqref{eq:synthesis_problem_unconst}, 
i.e., the synthesis formulation of the dequantization problem 
while the CP (Alg.\,\ref{alg:CP_dequantization}) approximates \eqref{eq:analysis_problem_unconst},
i.e., the analysis formulation.

The implementation was done in MATLAB 2019b 
and relies on the LTFAT toolbox \cite{LTFAT} in computing the signal synthesis and analysis.
The source codes are available at \url{http://www.utko.feec.vutbr.cz/~rajmic/software/sparse_dequant.zip}.

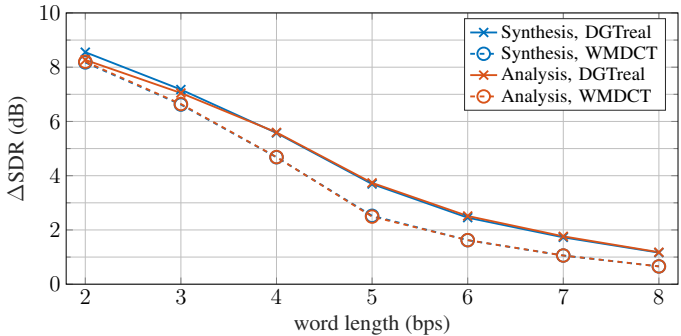
\begin{figure}[t]%
\definecolor{mycolor1}{rgb}{0.00000,0.44700,0.74100}%
\definecolor{mycolor2}{rgb}{0.85000,0.32500,0.09800}%
\begin{tikzpicture}[scale=0.59]

\begin{axis}[%
width=5.372in,%4.521in,
height=2.4in,%3.566in,
at={(0.758in,0.481in)},
scale only axis,
xmin=1.8,
xmax=8.2,
xlabel style={font=\Large\color{white!15!black}},
xlabel={word length (bps)},
xtick = {2, 3, 4, 5, 6, 7, 8},
xticklabel style={font=\Large},
minor y tick num=1,
grid=both,
ymin=0,
ymax=10,
ylabel style={font=\Large\color{white!15!black}},
ylabel={\dsdr{} (dB)},
yticklabel style={font=\Large},
axis background/.style={fill=white},
xmajorgrids,
ymajorgrids,
legend style={legend cell align=left, align=left, draw=white!15!black, font=\large},
every axis plot/.append style={very thick}
]
\addplot [color=mycolor1, mark=x, mark size=4pt, mark options={solid, mycolor1}]
  table[row sep=crcr]{%
2	8.5533221105798\\
3	7.17605420178673\\
4	5.57494197467239\\
5	3.69251769367675\\
6	2.4531881835826\\
7	1.72711867200197\\
8	1.16372304768408\\
};
\addlegendentry{Synthesis, DGTreal}

\addplot [color=mycolor1, dashed, mark=o, mark size=4pt, mark options={solid, mycolor1}]
  table[row sep=crcr]{%
2	8.17421149359494\\
3	6.62095289602473\\
4	4.67972294603438\\
5	2.52182759601637\\
6	1.62234767192683\\
7	1.05686184703308\\
8	0.655948526065191\\
};
\addlegendentry{Synthesis, WMDCT}

\addplot [color=mycolor2, mark=x, mark size=4pt, mark options={solid, mycolor2}]
  table[row sep=crcr]{%
2	8.26921725250611\\
3	7.04859287151692\\
4	5.59602958594475\\
5	3.73564328284118\\
6	2.51427280690092\\
7	1.76431592608647\\
8	1.1790679387983\\
};
\addlegendentry{Analysis, DGTreal}

\addplot [color=mycolor2, dashed, mark=o, mark size=4pt, mark options={solid, mycolor2}]
  table[row sep=crcr]{%
2	8.20279936713916\\
3	6.64303173793878\\
4	4.6864916567277\\
5	2.49156131303261\\
6	1.62457370678688\\
7	1.0575825877189\\
8	0.656512547642218\\
};
\addlegendentry{Analysis, WMDCT}

\end{axis}
\end{tikzpicture}%
\caption{Average performance in terms of \dsdr{}.}%
\label{fig:dSDR}%
\end{figure}

\subsection{Evaluation Using Signal-to-Distortion Ratio (SDR)}

The similarity of the restored signal to the original is evaluated using the \dsdr{}, 
which expresses the SDR improvement and is computed as the difference between the restored and the quantized signal according to:
\begin{equation}
\dsdr = \text{SDR}(\x, \hat{\x}) - \text{SDR}(\x, \xq).
\label{eq:dsdr}
\end{equation}
The SDR for two signals $\u$ and $\vv$ is computed as
\begin{equation}
\text{SDR}(\u, \vv) = 10\log_{10}\frac{\norm{\u}^2_2}{\norm{\u-\vv}^2_2}.
\label{eq:sdr}
\end{equation}

\begin{table*}[b]
\centering
\caption{\rule{0pt}{17pt}Parameter values used in the proximal algorithms.}
\label{tab:parameters}
\renewcommand{\arraystretch}{1.2}
\begin{tabular}{|c|c||c|c|c|c|c|c|c|}
\hline
\multirow{2}{*}{\textbf{Algorithm and parameter}} & \multirow{2}{*}{\textbf{Frame}} & \multicolumn{7}{c|}{\textbf{Word length}}                                     \\ \cline{3-9} 
                                         &                        & 2      & 3      & 4      & 5       & 6        & 7        & 8         \\ \hline\hline
\multirow{2}{*}{\begin{tabular}[c]{@{}c@{}}Douglas--Rachford\\$\gamma$\end{tabular}} &
  DGTreal &
  \multicolumn{1}{l|}{0.0073} &
  \multicolumn{1}{l|}{0.0040} &
  \multicolumn{1}{l|}{0.0015} &
  \multicolumn{1}{l|}{0.00025} &
  \multicolumn{1}{l|}{0.000049} &
  \multicolumn{1}{l|}{0.000017} &
  \multicolumn{1}{l|}{0.0000066} \\ \cline{2-9} 
                                         & WMDCT                  & 0.0204 & 0.0123 & 0.0055 & 0.00035 & 0.000084 & 0.000028 & 0.0000099 \\ \hline
\multirow{2}{*}{\begin{tabular}[c]{@{}c@{}}Chambolle--Pock\\$\zeta$\end{tabular}} &
  DGTreal &
  0.0055 &
  0.0031 &
  0.0013 &
  0.00017 &
  0.000041 &
  0.000015 &
  0.0000057 \\ \cline{2-9} 
                                         & WMDCT                  & 0.0213 & 0.0110 & 0.0053 & 0.00023 & 0.000066 & 0.000022 & 0.0000075 \\ \hline
\end{tabular}

\end{table*}

\noindent
The average \dsdr{} values for both proposed algorithms, and the DGT or the WMDCT as a transform, are presented in Fig.~\ref{fig:dSDR}.
According to the \dsdr{} results, using the DGT instead of the WMDCT leads to a better restoration quality by up to 1\,dB for both proposed algorithms.
On the other hand, no significant difference in the quality of restoration between the synthesis and the analysis model has been found;
The DR algorithm performs marginally better in the case of very harsh quantization, i.e., 2 and 3 bps.

\subsection{Evaluation Using PESQL}

Since \dsdr{} indicates only the physical similarity between two signals with no psychoacoustics involved, 
the quality of restoration was also evaluated using the Perceptual Evaluation of Speech Quality (PESQ), 
specifically, using the implementation provided in \cite{Loizou2017:Speech.Enhancement},
which is the same evaluator that was used in
\cite{BrauerGerkmannLorenz2016:Sparse.reconstruction.of.quantized.speech.signals}.
The output of the PESQL is the Mean Opinion Scores (MOS), which covers a scale of 1 (bad) to 5 (excellent).

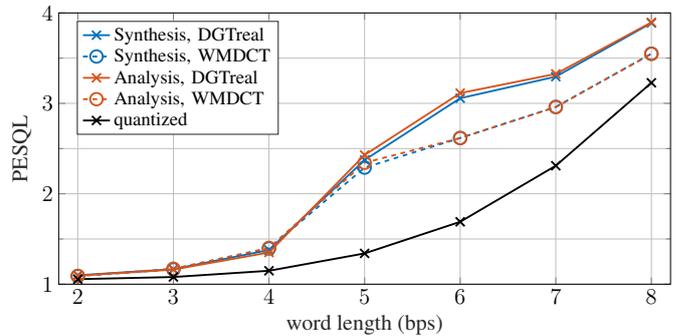
\begin{figure}[t]%
\definecolor{mycolor1}{rgb}{0.00000,0.44700,0.74100}%
\definecolor{mycolor2}{rgb}{0.85000,0.32500,0.09800}%
\begin{tikzpicture}[scale=0.59]%, show background rectangle]

\begin{axis}[%
width=5.372in,%4.521in,
height=2.4in,%3.566in,
at={(0.758in,0.481in)},
scale only axis,
xmin=1.8,
xmax=8.2,
xlabel style={font=\Large\color{white!15!black}},
xlabel={word length (bps)},
xtick = {2, 3, 4, 5, 6, 7, 8},
xticklabel style={font=\Large},
ymin=1,
ymax=4,
minor y tick num=1,
grid=both,
ylabel style={font=\Large\color{white!15!black}},
ylabel={PESQL},
yticklabel style={font=\Large},
ytick = {1, 2, 3, 4},
axis background/.style={fill=white},
xmajorgrids,
ymajorgrids,
legend style={at={(0.03,0.97)}, anchor=north west, legend cell align=left, align=left, draw=white!15!black, font=\large},
every axis plot/.append style={very thick}
]
\addplot [color=mycolor1, mark=x, mark size=4pt, mark options={solid, mycolor1}]
  table[row sep=crcr]{%
2	1.09857105718601\\
3	1.1662422320396\\
4	1.37798094059102\\
5	2.37393420427557\\
6	3.05663822665309\\
7	3.29542405748645\\
8	3.88873603937559\\
};
\addlegendentry{Synthesis, DGTreal}

\addplot [color=mycolor1, dashed, mark=o, mark size=4pt, mark options={solid, mycolor1}]
  table[row sep=crcr]{%
2	1.08985248014904\\
3	1.16345134198831\\
4	1.39535270043576\\
5	2.28954970825279\\
6	2.61682561158132\\
7	2.96049972498573\\
8	3.54930544955127\\
};
\addlegendentry{Synthesis, WMDCT}

\addplot [color=mycolor2, mark=x, mark size=4pt, mark options={solid, mycolor2}]
  table[row sep=crcr]{%
2	1.09339769496682\\
3	1.16201384357369\\
4	1.35076827038863\\
5	2.42908646844918\\
6	3.11315454817588\\
7	3.32594052890147\\
8	3.89591724674375\\
};
\addlegendentry{Analysis, DGTreal}

\addplot [color=mycolor2, dashed, mark=o, mark size=4pt, mark options={solid, mycolor2}]
  table[row sep=crcr]{%
2	1.09210469838856\\
3	1.17117109904043\\
4	1.40117457825015\\
5	2.34326015431495\\
6	2.61604527669779\\
7	2.95867635281346\\
8	3.54721224010981\\
};
\addlegendentry{Analysis, WMDCT}

\addplot [color=black, mark=x, mark size=4pt, mark options={solid, black}]
  table[row sep=crcr]{%
2	1.05553379191658\\
3	1.07939439576875\\
4	1.14851143640394\\
5	1.34053784461497\\
6	1.69009438434558\\
7	2.31035569897513\\
8	3.22759408681571\\
};
\addlegendentry{quantized}

\end{axis}
\end{tikzpicture}%
\caption{Average performance in terms of PESQL.}%
\label{fig:PESQL}%
\end{figure}

The PESQL results of the dequantization are shown in Fig.\,\ref{fig:PESQL}.
Along with the average PESQL results of the restored signals, the PESQL values of the quantized signals are also plotted in the figure.
Both proposed algorithms with either the DGT or the WMDCT performed very similarly for smaller word lengths (specifically 2, 3 and 4 bps).
For bigger word lengths (6, 7 and 8 bps) there is a more significant difference between the WMDCT and the DGT, 
suggesting that the DGT is more suitable for this type of restoration task. 
Also, the PESQL values suggest a slight advantage of using the analysis model, even though the difference in PESQL is small.

\subsection{Experiment Setup and Parameter Fine-Tuning}

To obtain the best possible results of dequantization, a careful selection of the parameters of the proximal algorithms must be made, 
specifically, the parameter $\gamma$ in the DR algorithm, which is used as a threshold for the soft thresholding step.
If $\gamma$ is too big, most of the time-frequency coefficients are pushed to zero and the algorithm does not converge properly.
On the other hand, a very small $\gamma$ causes the algorithm to converge very slowly. 
Similar behavior is observed with the parameter $\zeta$ used in the Chambolle--Pock algorithm.

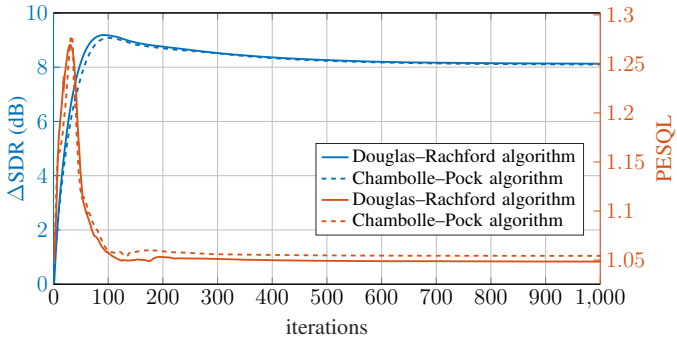
\begin{figure}[t]%
\definecolor{mycolor1}{rgb}{0.00000,0.44700,0.74100}%
\definecolor{mycolor2}{rgb}{0.85000,0.32500,0.09800}%
\begin{tikzpicture}[scale=0.59]
\begin{axis}[%
width=4.8in,%4.521in,
height=2.4in,%3.566in,
at={(0.503in,0.455in)},
scale only axis,
xmin=0,
xmax=1000,
xticklabel style={font=\Large},
xlabel style={font=\Large\color{white!15!black}},
xlabel={iterations},
separate axis lines,
every outer y axis line/.append style={mycolor1},
every y tick label/.append style={font=\Large\color{mycolor1}},
every y tick/.append style={mycolor1},
ymin=0,
ymax=10,
ylabel shift = -8pt,
axis y line*=left,
ylabel style={font=\Large\color{mycolor1}},
ylabel={$\Delta$SDR (dB)},
yticklabel pos=left,
xmajorgrids,
ymajorgrids,
every axis plot/.append style={very thick}
]
\addplot [color=mycolor1] table {figz/DR_dSDR.dat};
\label{dSDR_course_DR}
\addplot [color=mycolor1,dashed] table {figz/CP_dSDR.dat};
\label{dSDR_course_CP}
\end{axis}
\begin{axis}[%
width=4.8in,%4.521in,
height=2.4in,%3.566in,
at={(0.503in,0.455in)},
scale only axis,
xmin=0,
xmax=1000,
xmajorticks=false,
separate axis lines,
every outer y axis line/.append style={mycolor2},
every y tick label/.append style={font=\Large\color{mycolor2}},
every y tick/.append style={mycolor2},
axis y line*=right,
ylabel style={font=\Large\color{mycolor2}},
ylabel={PESQL},
yticklabel pos=right,
legend style={at={(0.97,0.18)}, anchor=south east, legend cell align=left, align=left, draw=white!15!black, font=\large},
every axis plot/.append style={very thick}
]
\addlegendimage{/pgfplots/refstyle=dSDR_course_DR}\addlegendentry{Douglas--Rachford algorithm}
\addlegendimage{/pgfplots/refstyle=dSDR_course_CP}\addlegendentry{Chambolle--Pock algorithm}
\addplot [color=mycolor2] table {figz/DR_PESQL.dat};
\addlegendentry{Douglas--Rachford algorithm}
\addplot [color=mycolor2,dashed] table {figz/CP_PESQL.dat};
\addlegendentry{Chambolle--Pock algorithm}
\legend{Douglas--Rachford algorithm, Chambolle--Pock algorithm, Douglas--Rachford algorithm, Chambolle--Pock algorithm}
\end{axis}
\end{tikzpicture}%
\caption{
Development of \dsdr{} values and PESQL through iterations for the testing signal \qm{S\_01\_01.wav} dequantized from 2 bps quantization for both synthesis (Douglas--Rachford) and analysis (Chambolle--Pock) model.
For this example, the real-valued DGT was selected as the default transform. 
}%
\label{fig:dSDR_PESQL_development}%
\end{figure}

The courses of the \dsdr{} and PESQL values through iterations for both algorithms are presented in Fig.~\ref{fig:dSDR_PESQL_development}. 
One can notice a fast gain in the first couple of iterations followed by a~slight drop, after which the \dsdr{} value stabilizes.
% on a certain value. 
%This behavior is most-likely explained as that the processed signals (especially speech signals) are not as sparse as the algorithm would need,
The most likely explanation for this behavior is that the processed signals (especially speech signals) are not as sparse as the algorithm would need,
%for example some music signals,
and after a certain number of iterations the algorithm continues to further \qm{sparsify} the restored signal,
making it less similar to the original one because it has lots of zeros on the positions of the lowest quantization levels,
which is also confirmed by the observations of the restored signals.

This is the reason why we prefer to terminate the algorithms at the SDR peaks. 
Since computational time is not crucial in this application, we set the respective parameters $\gamma,\zeta$ for the DR and the CP algorithms, respectively, to terminate the algorithms after approximately 100 iterations.
The specific values used for testing
%can be found
are
in Table\,\ref{tab:parameters}.
Apart from the parameters, we also set the maximum number of iterations for both algorithms to 400 and the minimum number of iterations to 50.

The other parameters were set to $\lambda = 1$ for the Douglas--Rachford and $\sigma = 1/\zeta, \rho = 1$ for the Chambolle--Pock.

As the sparsity-promoting transforms, the real-valued discrete Gabor transform (DGT) and the Windowed Modified Discrete Cosine Transform (WMDCT) were used. 
For both transforms, a 1024 samples long Hann window with 1024 frequency channels was used. 
The redundancy of the DGT was set to 4, i.e., the window overlap was 75\,\%.

% ---------------------------------------------------
% Conclusion
% ---------------------------------------------------

\section{Conclusion}
In this paper, two sparsity-based approaches to speech dequantization have been proposed,
using the synthesis and the analysis models of the signal.
The synthesis variant of the optimization problem is numerically solved with the Douglas--Rachford algorithm and the analysis variant with the Chambolle--Pock algorithm.
Both methods perform similar to each other and they lead to a significant improvement in the quality of restored audio in terms of \dsdr{} and PESQL.
The analysis model seems to perform slightly better according to the PESQL values. 
Both algorithms have also been tested using the WMDCT and the DGT signal transforms, %and the results play in the favor of using DGT 
and better results---according to both the \dsdr{} and PESQL---have been achieved using the DGT.

\vspace{1em}

% ---------------------------------------------------
% References
% ---------------------------------------------------

%\bibliographystyle{IEEEtran}

{
\bibliographystyle{IEEEtran}
\inputencoding{cp1250}
\bibliography{IEEEabrv,literatura}
}

% ---------------------------------------------------
% End of Document
% ---------------------------------------------------

\end{document}